\documentclass[amssymb,amsmath,aps,showpacs,floatfix,nofootinbib,showpacs,12pt,prd]{revtex4}

\usepackage{graphicx}
\usepackage{color}
\usepackage{comment}
\usepackage{ulem}

\begin{document}

\title{Leptonic dark matter annihilation in the evolving universe:
constraints and implications}

\author{Qiang Yuan$^{a,f}$}
\author{Bin Yue$^{b,f}$}
\author{Xiao-Jun Bi$^{c,a}$}
\author{Xuelei Chen$^{b,c}$}
\author{Xinmin Zhang$^{d,e}$}

\affiliation{$^a$Key Laboratory of Particle Astrophysics,
Institute of High Energy Physics, Chinese Academy of Science,
Beijing 100049, P.R.China}

\affiliation{$^b$National Astronomical Observatories, Chinese Academy
of Sciences, Beijing 100012, P.R.China}

\affiliation{$^c$Center for High Energy Physics, Peking
University, Beijing 100871, P.R.China}

\affiliation{$^d$Theoretical Physics Division, Institute of High
Energy Physics, Chinese Academy of Science, Beijing 100049, P.R.China}

\affiliation{$^e$Theoretical Physics Center for Science Facilities
(TPCSF), Chinese Academy of Science, Beijing 100049, P.R.China}

\affiliation{$^f$Graduate University of Chinese Academy of
Sciences, Beijing 100049, P.R.China}

\begin{abstract}

The cosmic electron and positron excesses have been explained as
possible dark matter (DM) annihilation products. In this work we
investigate the possible effects of such a DM annihilation scenario
during the evolution history of the Universe. We first calculate the
extragalactic $\gamma$-ray background (EGRB), which is produced
through the final state radiation of DM annihilation to charged
leptons and the inverse Compton scattering between
electrons/positrons and the cosmic microwave background. The DM halo
profile and the minimal halo mass, which are not yet well determined
from the current N-body simulations, are constrained by the EGRB
data from EGRET and Fermi telescopes. Then we discuss the impact of
such leptonic DM models on cosmic evolution, such as the
reionization and heating of intergalactic medium, neutral Hydrogen
21 cm signal and suppression of structure formation. We show that
the impact on the Hydrogen 21 cm signal might show interesting
signatures of DM annihilation, but the influence on star formation is
not remarkable. Future observations of the 21 cm signals could be
used to place new constraints on the properties of DM.

\end{abstract}

%95.35.+d: Dark matter
%95.85.Pw: gamma-rays
%98.58.Ge: HI regions and 21-cm lines; diffuse, translucent, and high-velocity clouds

\maketitle

\section{Introduction}

Recently, observations of the cosmic ray (CR) positron fraction by
the PAMELA experiment \cite{2009Natur.458..607A}, the total electron
and positron spectra by the ATIC \cite{2008Natur.456..362C}
experiment, the PPB-BETS \cite{2008arXiv0809.0760T} experiment, the
HESS \cite{2008PhRvL.101z1104A,2009A&A...508..561A} experiment and
the Fermi \cite{2009PhRvL.102r1101A} experiment all show interesting
excesses when compared with the expectations of the conventional
astrophysical background. Especially the PAMELA data confirm the
hint of excess of positron fraction previously discovered by HEAT
\cite{1997ApJ...482L.191B,2004PhRvL..93x1102B} and AMS
\cite{2007PhLB..646..145A}. Many theoretical models have been
proposed to explain these phenomena in past years, among which a
class of leptonic dark matter (DM) annihilation models is of
particular interest (e.g., \cite{2008PhRvD..78j3520B,
2009NuPhB.813....1C} or see the review \cite{2009MPLA...24.2139H}).

It was soon recognized that if DM annihilation can occur near the
solar location to contribute to the locally observed CR leptons, it
should occur everywhere in the Universe as long as DM exists.
Shortly after the proposal of using DM annihilation to account for
the data, the $\gamma$-ray and radio emissions from the Galactic
center
\cite{2009JCAP...03..009B,2009PhRvD..80b3007Z,2009PhRvD..79h1303B},
the Milky Way halo \cite{2009ApJ...699L..59B,2009NuPhB.821..399C,
2009PhRvD..80d3525R,2009arXiv0908.1236Z,2009arXiv0912.0663C}, the
dwarf galaxies
\cite{2009MNRAS.399.2033P,2009PhRvD..80b3506E,2010AdAst2010E..45K},
the galaxy clusters \cite{2009PhRvD..80b3005J,2009PhRvL.103r1302P}
and the early Universe
\cite{2008PhRvL.101z1301K,2009PhRvD..80b3517K, 2009JCAP...07..020P}
were studied as possible constraints and/or future probes of the DM
models. These studies have shown that the proposed DM scenarios are
either excluded or at least strongly constrained if the DM halo
properties derived from current numerical simulations are adopted.
However, we should bear in mind that there are still large
uncertainties in our understanding of the DM halo structure,
especially for the small scales which are not yet probed by the
current simulations due to the limited resolution. In the present
work, we adopt the leptonic DM scenario implied from the CR data,
and investigate what kind of constraint could be placed on the
properties of dark matter halos \cite{2009PhRvD..80j3502B,
2009PhRvL.103r1302P}. This exercise is useful in two ways. On one
hand with the improvement of computing power and astrophysical
observational precision, we will achieve better understanding of the
DM halo, if in the future our DM-derived constraints are in conflict
with these simulations or observations, then it helps to falsify the
DM model considered here. On the other hand, it also helps to
illustrate the possible model uncertainties in explaining the CR
observations.

In this work we shall focus on the DM annihilation during the
evolution history of the Universe, mainly after the epoch of cosmic
recombination. The DM annihilation can produce electrons/positrons,
which can scatter with the cosmic microwave background (CMB) photons
to produce $\gamma$-rays and contribute to the extragalactic
$\gamma$-ray background (EGRB). The final state radiation (FSR) from
DM annihilation into charged leptons (generated directly from the
external legs of the Feynman diagram) can also contribute to the
EGRB. Using the observations on the EGRB by EGRET
\cite{1998ApJ...494..523S} and most recently by Fermi
\cite{2010PhRvL.104j1101A}, one can test the model configurations
and constrain the model parameters
\cite{2009PhRvD..80b3517K,2009JCAP...07..020P}. In addition, the
energy of the final state particles from DM annihilation will be
partially deposited in the baryonic gas, causing ionization and
heating. This effect can be imprinted on the CMB anisotropy
\cite{2004PhRvD..70d3502C,2005PhRvD..72b3508P,
2006PhRvD..74j3519Z,2006MNRAS.369.1719M,2009PhRvD..80b3505G,
2009arXiv0907.3985K}, the optical depth of CMB photons
\cite{2009A&A...505..999H,2009JCAP...10..009C}, and the 21 cm signal
from the neutral hydrogen during the dark age
\cite{2006PhRvD..74j3502F,2008ApJ...679L..65C,2009PhRvD..80d3529N}.
Furthermore, the heating also increases the Jeans mass, suppressing
the accretion of gas into minihalos, and enhancing the
photoevaporation of gas in minihalos \cite{2007MNRAS.375.1399R}.
These effects might affect the early star formation history of the
Universe.

Different from previous studies, in this work we will normalize the
particle physics parameters of DM particles (including mass, cross
section, branching ratios etc.) to the recent observed CR lepton
data \cite{2010PhRvD..81b3516L}\footnote{We adopt the parameters
fitted using fixed background approach given in Ref.
\cite{2010PhRvD..81b3516L} (Table I).}, then calculate the EGRB and
derive the constraints on the DM halo profile and the minimal halo
mass using EGRET and Fermi data. Then we discuss the possible
effects of DM annihilation on the ionization and heating of the
intergalactic gas, the 21 cm signal and the early structure
formation. Throughout this paper we will adopt the cosmological
parameters from WMAP 5-yr results combined with the baryon acoustic
oscillation and Type Ia supernova data: $\Omega_\chi=0.228$,
$\Omega_\Lambda=0.726$, $\Omega_b=0.046$, $h=0.705$
,$\sigma_8=0.812$ and $n_s=0.96$ \cite{2009ApJS..180..330K}. The
uncertainties of the cosmological parameters would affect the
detailed result of the clumpiness factor of DM. However, compared
with the large uncertainties of other aspects in the study, e.g.,
the concentration model and structure formation history of DM, the
uncertainties of cosmological parameters would play a negligible
role.

\section{Extragalactic $\gamma$-ray background}

The $\gamma$-ray emission produced in the leptonic DM annihilation
generally include two components: the inverse Compton (IC) component
through interactions between electrons/positrons and the
cosmological radiation field such as the CMB, and the direct FSR
associated with the annihilation process itself. The $\gamma$-ray
flux observed at redshift $z$ is obtained by integrating the
emissivity through evolution history of the universe
\cite{2002PhRvD..66l3502U,2009A&A...505..999H},
\begin{equation}
\phi(E,z)=\frac{c(1+z)^2}{4\pi}
\frac{\Omega_\chi^2\rho_c^2\langle\sigma v\rangle}{2m_\chi^2}
\int_z^{\infty}{\rm
d}z'\frac{(1+z')^3[\Delta^2(z')+1]}{H(z')}\frac{{\rm d}N}{{\rm d}E'}
\exp\left[-\tau(z;z',E')\right], \label{phi}
\end{equation}
where $m_\chi$ is the mass of DM particle, $\langle\sigma v\rangle$
is the velocity weighted annihilation cross section,
$\rho_c=3H_0^2/8\pi G$ is the critical density of the Universe at
present, $H(z)=H_0\sqrt{
(\Omega_\chi+\Omega_b)(1+z)^3+\Omega_\Lambda}$ is the Hubble
parameter. Here $E'\equiv E(1+z')/(1+z)$, and $\frac{{\rm d}N}{{\rm
d}E'}=\left.\frac{{\rm d}N}{{\rm d}E'}\right| _{\rm
IC}+\left.\frac{{\rm d}N}{{\rm d}E'}\right|_{\rm FSR}$ is the
$\gamma$-ray generation multiplicity at redshift $z'$ for one pair
of annihilating DM particles. Note that the cascade contribution
from pair production of the high energy photons interacting with CMB 
photons is not included. This treatment is reasonable due to the following
two arguments. On one hand for the DM models considered in this work (see 
below Table \ref{table:sv}), the energy fraction that goes into FSR 
photons, which should be easier to reach the pair production threshold,
is much smaller than that which goes into electrons/positrons.
On the other hand, because of the large clumpiness factor the dominant 
contribution to the EGRB for most of the redshift range interested here 
(e.g., $z\lesssim 100$) comes from the DM annihilation at relatively low 
redshifts, where the pair production is also not important.
The optical depth $\tau(z;z',E')$ of $\gamma$-ray photon with energy
$E'$, propagating from $z'$ to $z$ can be obtained as
\begin{equation}
\tau(z;z',E')=c\int_z^{z'}{\rm
d}z''\frac{\alpha(E'',z'')}{H(z'')(1+z'')},
\end{equation}
where $E''=E'(1+z'')/(1+z')$, $\alpha(E,z)$ is the absorption
coefficient. In this work we include the following interactions for
photon propagation: the photo-ionization, photon-nuclei pair
production, Compton scattering, photon-photon scattering and
photon-photon pair production. The absorption coefficients of these
processes are adopted from \cite{1989ApJ...344..551Z}. Note that for
$6\lesssim z\lesssim 1200$ the Universe is neutral, therefore the
Compton scattering is negligible; otherwise when the Universe is
ionized the photo-ionization is dropped. For redshift less than $6$
we further include the pair production process between $\gamma$-ray
photons and cosmic infrared background, using the optical depth of
the ``baseline'' model in \cite{2006ApJ...648..774S}. Finally
$\Delta^2(z)$ is a factor describing the clumpiness of the DM
structures. The detailed calculation of $\Delta^2(z)$ using
structure formation theory is presented in the Appendix.

The FSR photon spectrum for the $e^+e^-$ or $\mu^+\mu^-$ channel, when
$m_{\chi}\gg m_e,\,m_{\mu}$, is given by \cite{2005PhRvL..94m1301B}
\begin{equation}
\left.\frac{{\rm d}N}{{\rm d}x}\right|_{\rm FSR}^i=\frac{\alpha}{\pi}
\frac{1+(1-x)^2}{x}\log\left(\frac{s}{m_i^2}(1-x)\right),
\label{fsr_emu}
\end{equation}
where $i=e,\,\mu$, $\alpha\approx 1/137$ is the fine structure constant,
$s=4m_{\chi}^2$ and $x=E/m_{\chi}$. For the $\tau^+\tau^-$ channel,
there is an internal bremsstrahlung radiation
from the chain $\tau\rightarrow\pi^0\rightarrow\gamma$
in addition to the decay products. The total radiation is given by
\cite{2004PhRvD..70j3529F}
\begin{equation}
\left.\frac{{\rm d}N}{{\rm d}x}\right|_{\rm FSR}^{\tau}=x^{-1.31}
\left(6.94x-4.93x^2-0.51x^3\right)e^{-4.53x}.
\label{fsr_tau}
\end{equation}

The IC component of $\gamma$-ray photons is given by
\begin{equation}
\left.\frac{{\rm d}N}{{\rm d}E}\right|_{\rm IC}=\int {\rm d}\epsilon\,
n(\epsilon)\int {\rm d}E_e \frac{{\rm d}n}{{\rm d}E_e} \times
F_{\rm KN}(\epsilon,E_e,E),
\label{dNdEIC}
\end{equation}
where $n(\epsilon)$ is the photon number density of the background
radiation as a function of energy $\epsilon$,
and $\frac{{\rm d}n}{{\rm d}E_e}$ is the quasi-equilibrium
electron/positron energy spectrum from the DM annihilations. Following
\cite{2009JCAP...07..020P} we assume that the electrons/positrons
lose their energies instantaneously
after the production from DM annihilation,
\begin{equation}
\frac{{\rm d}n}{{\rm d}E_e}=\frac{1}{b(E_e,z)}\int_{E_e}^{m_\chi}
{\rm d}E_e'\frac{{\rm d}N_e}{{\rm d}E_e'},
\end{equation}
with the energy loss rate $b(E_e,z)\approx 2.67\times 10^{-17}(1+z)^4
\left(E_e/{\rm GeV}\right)^2$ GeV s$^{-1}$.
Finally the differential Klein-Nishina cross section
$F_{\rm KN}(\epsilon,E_e,E)$ in Eq.(\ref{dNdEIC}) is given by
\begin{equation}
F_{\rm KN}(\epsilon,E_e,E)=\frac{3\sigma_T}{4\gamma^2\epsilon}\left[2q
\ln{q}+(1+2q)(1-q)+\frac{(\Gamma q)^2(1-q)}{2(1+\Gamma q)}\right],
\end{equation}
with $\sigma_T$ the Thomson cross section, $\gamma$ the Lorentz factor
of electron, $\Gamma=4\epsilon\gamma/m_e$, and $q=\frac{E/E_e}
{\Gamma(1-E/E_e)}$. For $q<1/4\gamma^2$ or $q>1$ we set
$F_{\rm KN}(\epsilon,E_e,E)=0$.

\begin{figure}[!htb]
\begin{center}
\includegraphics[width=0.45\columnwidth]{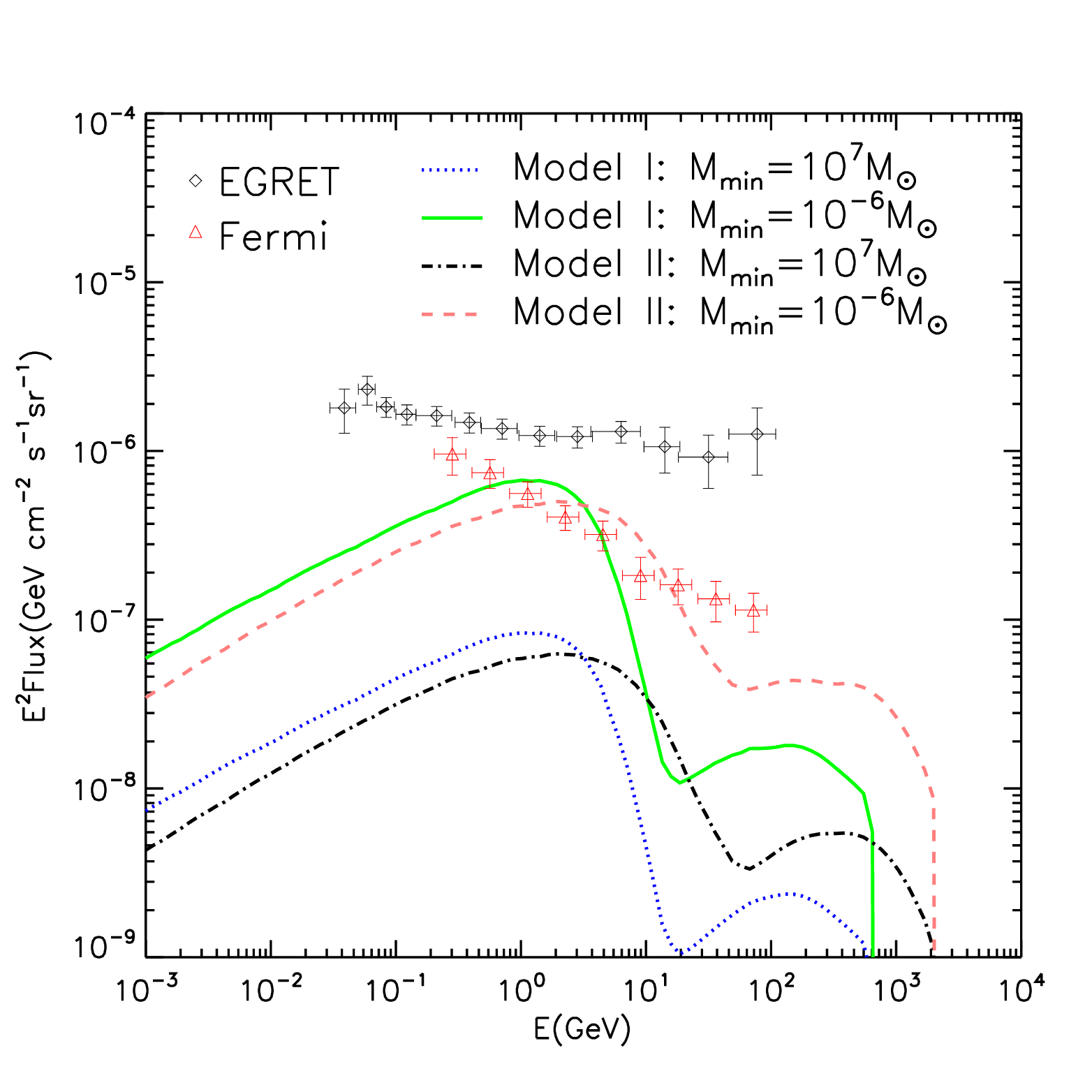}
\includegraphics[width=0.45\columnwidth]{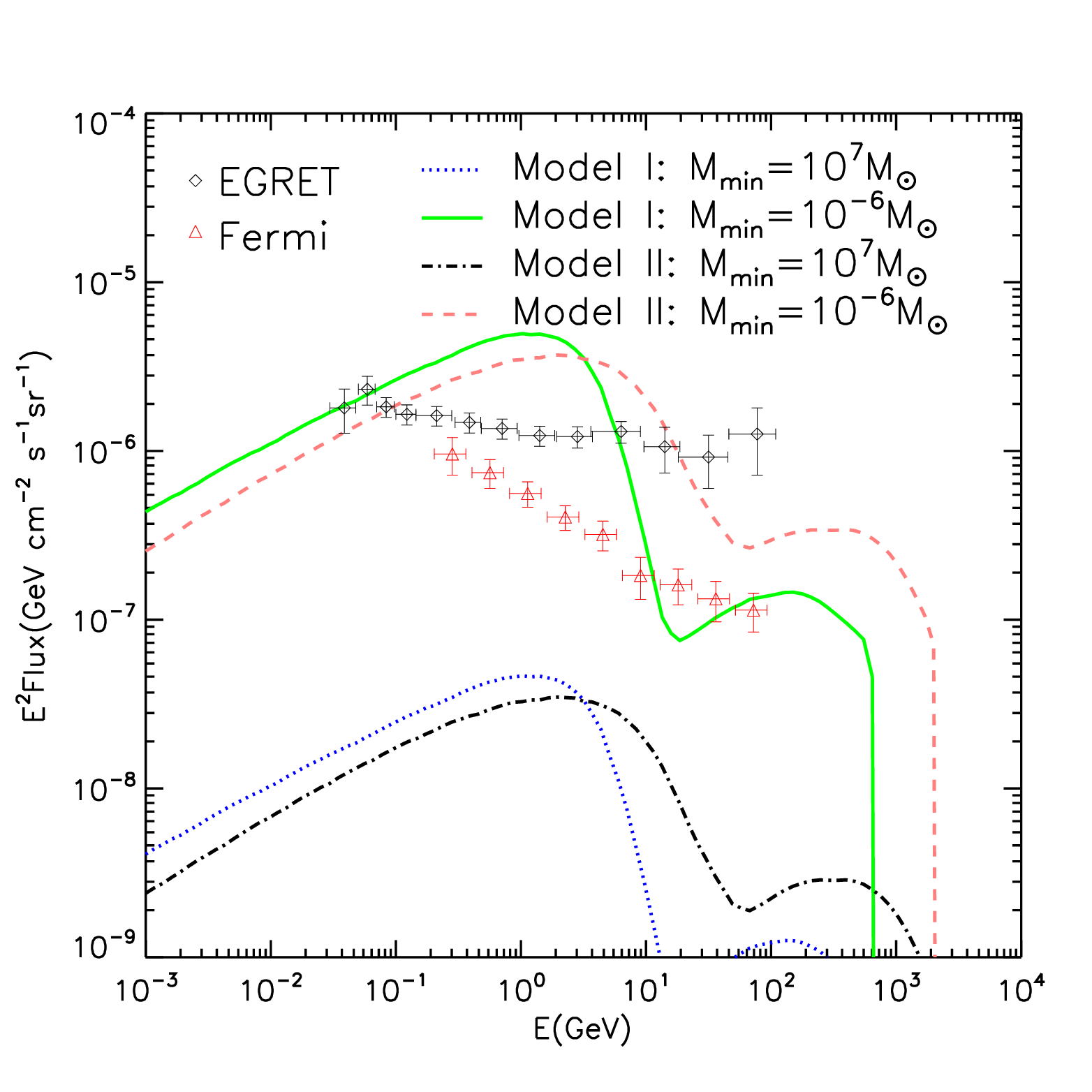}
\caption{EGRB from DM annihilation compared with the EGRET and Fermi data.
{\it Left} panel is for B01 concentration, and the {\it right} panel is
for M08 concentration model respectively. NFW halo profile is adopted.
\label{fig:eg}}
\end{center}
\end{figure}

The calculated EGRB intensity is plotted in Fig.~\ref{fig:eg}. Here
we adopt two sets of DM particle parameters, as compiled in Table
\ref{table:sv}, which are derived by a Markov Chain Monte Carlo
(MCMC) fitting of the PAMELA+ATIC (Model I) and PAMELA+Fermi+HESS
(Model II) data respectively \cite{2010PhRvD..81b3516L}. The cross
sections derived in \cite{2010PhRvD..81b3516L} are based on a local
DM density $\rho_{\odot}\approx0.25-0.27$ GeV cm$^{-3}$, which seems
to be smaller than the value derived in recent studies
\cite{2010JCAP...08..004C,2010arXiv1003.3101S}. Also, 
\cite{2010PhRvD..82b3531P} shows that the dynamic measurements might 
systematically under-estimate the local density. A
larger local density will result in a smaller cross section (by a
factor of $\sim2$) if we normalize the electron and positron fluxes
to the observational excesses. Since the present work is more or
less an order of magnitude study, we do not expect the change of
factor $2$ will affect the qualitative conclusions in this work. But
keep in mind that the signals discussed in the following should be
recognized as upper limits of such scenarios.

The {\it Left} and {\it right} panels are for two concentration
models of DM halos: Bullock et al. (B01, \cite{2001MNRAS.321..559B})
and Maccio et al. (M08, \cite{2008MNRAS.391.1940M}) respectively.
The expected EGRB intensity is a function of the DM halo parameters
$\gamma$ and $M_{\rm min}$ (see the Appendix). In this plot we adopt
$\gamma=1$ (NFW profile, \cite{1997ApJ...490..493N}), and varying
$M_{\rm min}$ from $10^7$ M$_{\odot}$ to $10^{-6}$ M$_{\odot}$,
corresponding to the observational lower limit of dwarf galaxies
and theoretical lower limit of the cold DM scenario. It is shown that
for B01 concentration the parameter settings with $M_{\rm
min}=10^{-6}$ M$_{\odot}$ are allowed by EGRET data but marginally
excluded by Fermi data. For M08 concentration model, the extension
of halos to very low mass range seems to over produce $\gamma$-rays
and conflict with both sets of data.

\begin{table}[htb]
\caption{Mass, annihilation cross section and branching ratios of DM
particles used in this work.}
\begin{tabular}{cccccc}
\hline \hline
  & $m_{\chi}$ & $\langle\sigma v\rangle$ & $B_e$ & $B_{\mu}$ & $B_{\tau}$ \\
  & (TeV) & ($10^{-23}$ cm$^3$s$^{-1}$) & & & \\
\hline
  Model I  & $0.71$ & $0.97$ & $0.8$ & $0.1$ & $0.1$ \\
  Model II & $2.21$ & $9.10$ & $0.0$ & $0.7$ & $0.3$ \\
  \hline
  \hline
\end{tabular}
\label{table:sv}
\end{table}

\begin{figure}[!htb]
\begin{center}
\includegraphics[width=0.45\columnwidth]{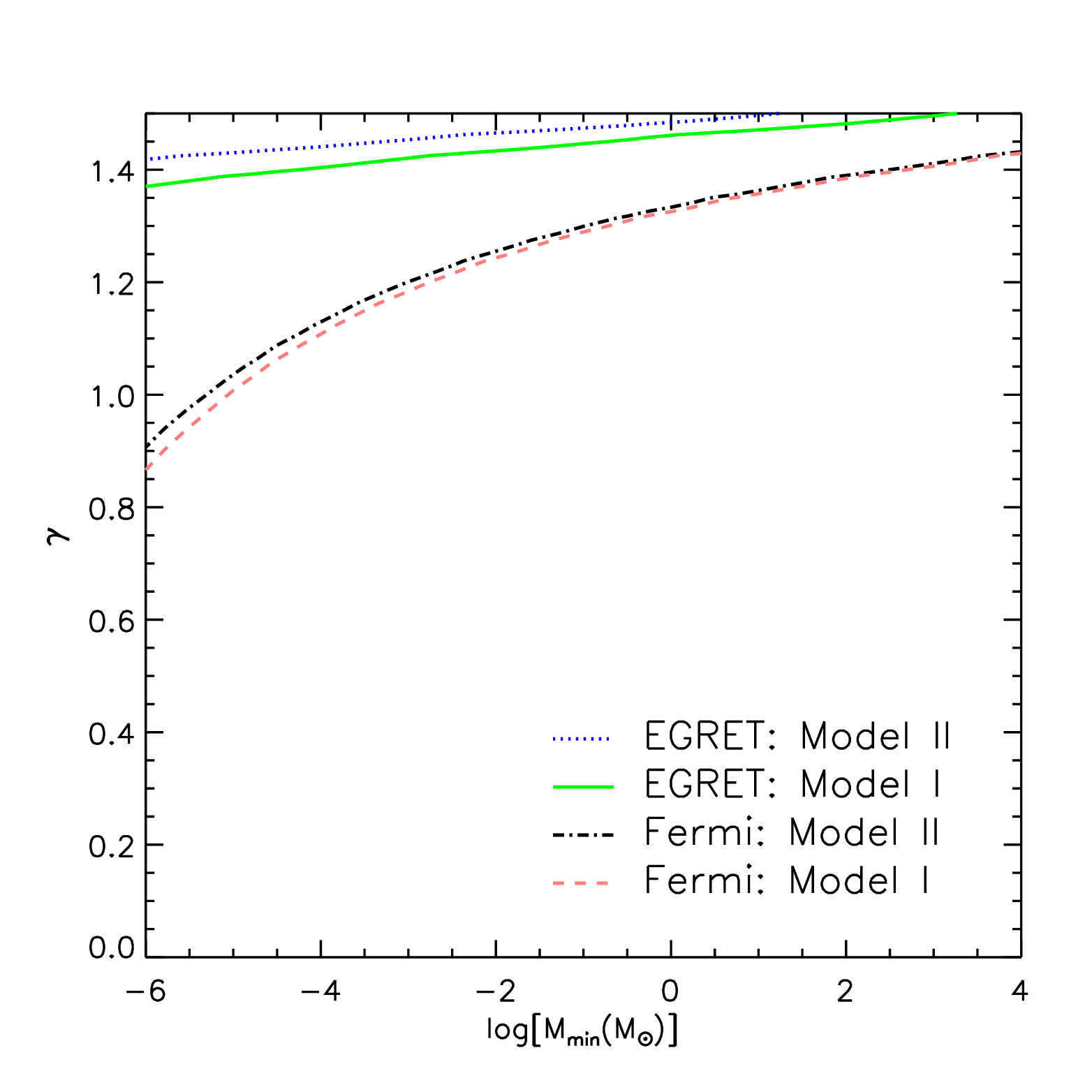}
\includegraphics[width=0.45\columnwidth]{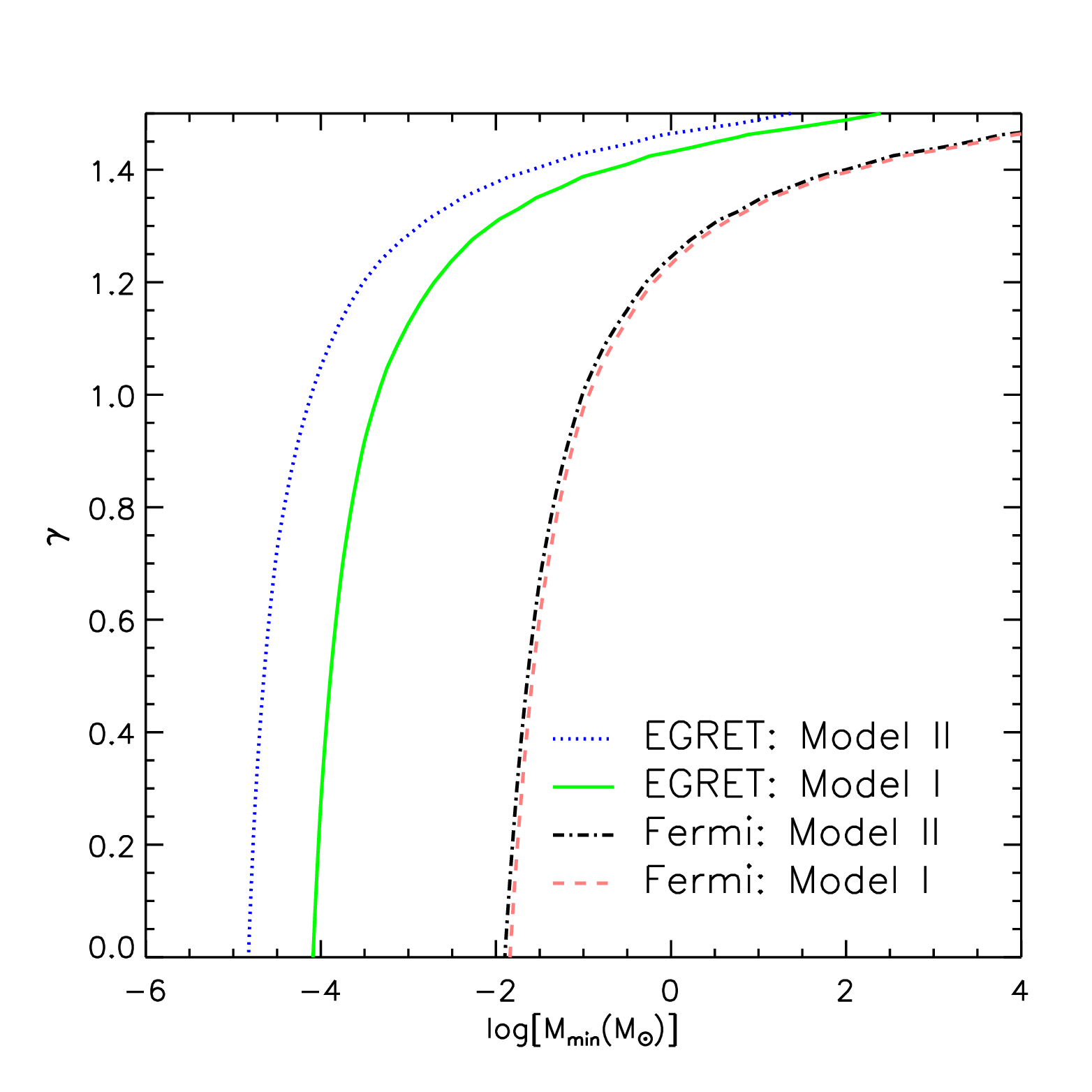}
\caption{Constraints on the DM halo parameters $M_{\rm min}$ and
$\gamma$ from EGRET and Fermi observations of EGRB. {\it Left}
panel is for B01 concentration, and the {\it right} panel is
for M08 concentration.
\label{fig:con}}
\end{center}
\end{figure}

We then scan the two dimensional parameter space $M_{\rm min}-\gamma$,
and calculate the expected EGRB for each setting of parameters.
By requring the calculated spectra not to exceed $2\sigma$
errorbars of the observational data, we derive the constraints
on these two parameters, as shown in Fig. \ref{fig:con}.
The parameter space to the top-left side of each curve is excluded.
The {\it left} panel is for B01 concentration model and the {\it right}
panel is for M08 concentration model. In each panel we plot
two groups of lines, corresponding to the EGRET and Fermi constraints
respectively. We can see that the constraints are sensitive to the
concentration models. The M08 concentration model were obtained by a
power-law fitting in the high mass range and then extrapolated to the
low mass range, while for B01 model the concentration of the low
mass haloes shows an asymptotic flattening. So for M08 model the
low mass haloes contribute more significantly to the $\gamma$-ray
intensity than that of B01 model, hence the constraint is
more stringent for M08 model when $M_{\rm min}$ is small.

From the results of the fitting, we find that the configuration of
standard cold DM haloes, e.g., the NFW profile ($\gamma=1$,
\cite{1997ApJ...490..493N}) and $M_{\rm min}$ down to $\sim 10^{-6}$
M$_{\odot}$, is constrained by the Fermi data. A warm DM scenario
with non-thermal production of DM particles seems to be safely
consistent with the data\footnote{See also the constraints
derived in the Galactic center \cite{2009PhRvD..80j3502B} and Virgo
cluster \cite{2009PhRvL.103r1302P}.}\cite{2001PhRvL..86..954L}.

\section{Implications}

\subsection{Cosmic reionization and heating}

Now we turn to the effects of DM annihilation on the cosmic reionization
and heating. The energy deposited in the intergalactic medium (IGM) from
DM annihilation can be expressed as \cite{2005PhRvD..72b3508P}
\begin{equation}
\epsilon_{\rm DM}(z)=f(z)\left(\frac{\langle\sigma v\rangle
\Omega_\chi^2\rho_c^2}{2m_\chi^2\, n_{\rm b0}}\right)(1+z)^3\cdot
2m_\chi,
\end{equation}
where $n_{\rm b0}$ is the present number density of baryon, and
function $f(z)$ represents the efficiency of energy transfer from DM
to the IGM. In \cite{2009PhRvD..80d3526S} the efficiency $f(z)$ is
calculated in detail and parameterized with fitting functions,
without including the DM clumpiness. However, it can not be directly
used with the existence of DM structures, because the mean free path
of photons for some energies (e.g., $\sim$GeV) is large enough that
the evolution of structures is necessary to be taken into account.
In this work we follow the method given in
\cite{2009A&A...505..999H} to calculate the energy deposition rate
$f(z)$. In short $f(z)$ is defined as \cite{2009A&A...505..999H}
\begin{equation}
f(z)\equiv\frac{\int{\rm d}E\,E\phi(E,z)\alpha(E,z)}{\langle\sigma
v\rangle \Omega_{\chi}^2\rho_c^2(1+z)^6/m_{\chi}},
\end{equation}
where $\phi(E,z)$ is given by Eq.~(\ref{phi}). The denominator of
the above equation represents the energy output of DM annihilation
without the clumpiness, and the numerator is the energy absorbed by
IGM. Note here the energy transfer from DM annihilation to IGM is
mainly due to the absorption of photons, no matter what the
annihilation final state is. For details of the results of $f(z)$
please refer to \cite{2009A&A...505..999H}. Our calculated results
are slightly different from that given in \cite{2009A&A...505..999H}
due to different branching ratios of DM annihilation.

In addition to the energy injection from DM annihilation, X-rays can
also heat the neutral IGM up to a high temperature
\cite{2004ApJ...602....1C,2006PhR...433..181F,2006ApJ...652..849F,
2008ApJ...684...18C}. We assume that the X-ray emissivity is
proportional to the star formation rate, which at high redshifts is
proportional to the differential increase of baryon collapse
fraction \cite{2008ApJ...684...18C} and is given by
\cite{2006MNRAS.371..867F}(see also \cite{2001ApJ...553..499O}),
\begin{equation}
\epsilon_X(z)=1.09\times10^{-31}f_X f_\star\left[\frac{\rho_{b,0}
(1+z)^3}{\rm M_\odot Mpc^{-3}}\right]\left | \frac{{\rm d}f_{\rm coll}}
{{\rm d}z}\right |(1+z)hE(z)\ [\rm eV\ cm^{-3}s^{-1}],
\end{equation}
where $E(z)=\sqrt{\Omega_m(1+z)^3+\Omega_{\Lambda}}$,
$f_X\sim 1.0$ is a renormalization factor, $f_\star$ is
star formation efficiency and $f_{coll}$ is the collapse fraction.
We take $f_{X}=1$ and $f_\star=10^{-3}$.

Given the energy injection rate, the evolution equations of the
ionization fractions and the IGM temperature are modified
through adding the following terms to the standard equations
\cite{2005PhRvD..72b3508P}
\begin{eqnarray}
-\delta\left(\frac{{\rm d}x_{\rm H}}{{\rm d}z}\right)&=&\frac{\epsilon_{\rm
DM}(z)+\epsilon_{\rm X}(z)}{13.6{\,\rm eV}}\frac{1-x_{\rm H}}{3(1+f_{\rm He})}
\frac{1}{H(z)(1+z)},\\
-\delta\left(\frac{{\rm d}x_{\rm He}}{{\rm d}z}\right)&=&\frac{\epsilon_{\rm
DM}(z)+\epsilon_{\rm X}(z)}{24.6{\,\rm eV}}\frac{1-x_{\rm He}}{3(1+f_{\rm He})}
\frac{1}{H(z)(1+z)},\\
-\delta\left(\frac{{\rm d}T_{\rm igm}}{{\rm d}z}\right)&=&\frac{2[\epsilon_{\rm
DM}(z)+\epsilon_{\rm X}(z)]}{3}\frac{1+2x_{\rm H}+f_{\rm He}(1+2x_{\rm He})}{3(1+f_{\rm He})}
\frac{1}{H(z)(1+z)},
\end{eqnarray}
where $f_{\rm He}\approx 0.083$ is the ratio of the number density of
helium to that of hydrogen for the Universe, with the helium mass
abundance $Y_{\rm He}=0.25$. The ionization fraction of species $A$ is
defined as $x_A\equiv\frac{n_{A^+}}{n_{A^+}+n_A}$.
These set of differential equations are solved using the RECFAST code
\cite{1999ApJ...523L...1S}.

\begin{figure}[!htb]
\begin{center}
\includegraphics[width=0.6\columnwidth]{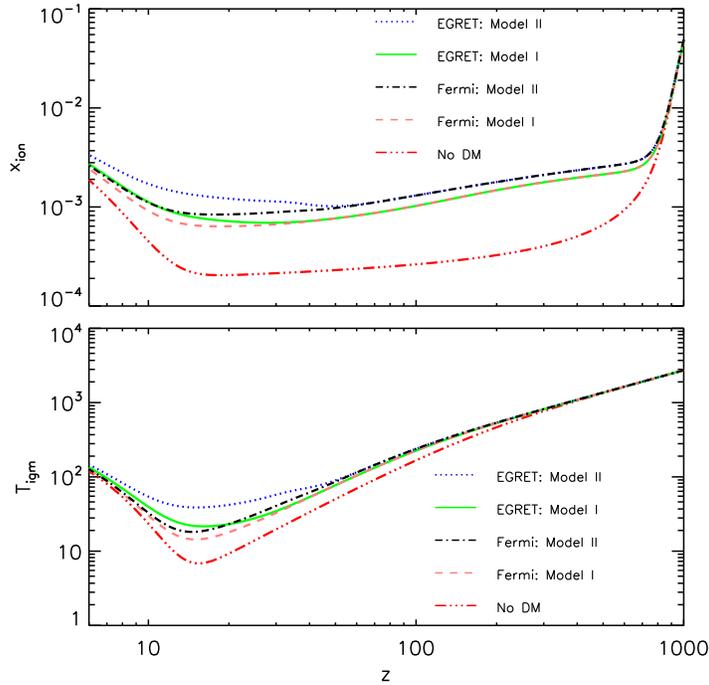}
\caption{The evolution of the cosmic ionization fraction ({\it upper}
panel) and the IGM temperature ({\it lower} panel), under the constraints
from EGRB.
\label{fig:xt}}
\end{center}
\end{figure}

In Fig. \ref{fig:xt} we show the changes of the total ionization fraction
$x_{\rm ion}=x_{\rm H}+f_{\rm He}x_{\rm He}$ and the IGM temperature
as functions of redshift, in the presence of DM annihilation. The four
curves correspond to the maximum clumpiness enhancement allowed by
EGRET or Fermi data, for the two DM particle models. Actually, for
different combinations of the parameters
$\gamma$ and $M_{\rm min}$ which lies on each of the curve in
Fig.~\ref{fig:con} the EGRB is of the same level. However, the impact
on the ionization and heating histories could be slightly different.
The parameters we chose in this study are compiled in Table
\ref{table:halo}. Here we adopt the B01 concentration model and try
to include as many low mass haloes as possible (i.e., the left
end of each curve in Fig. \ref{fig:con}).

\begin{table}[htb]
\caption{The halo parameters used to calculate the reionization and
heating processes. B01 concentration is adopted.}
\begin{tabular}{ccccccc}
\hline \hline
  & Model I (EGRET) & Model II (EGRET) & Model I (Fermi) & Model II (Fermi)\\
\hline
  $\gamma$      & $1.37$    & $1.42$    & $0.85$      & $0.90$    \\
  $M_{\rm min}$(M$_{\odot}$) & $10^{-6}$ & $10^{-6}$ & $10^{-6}$ & $10^{-6}$ \\
  \hline
  \hline
\end{tabular}
\label{table:halo}
\end{table}

We can see that at $z\approx 800$ the effects of DM annihilation on
the recombination process begin to appear, which lead to a higher
plateau of the cosmic ionization fraction. For $z\lesssim 50$ the
clumpiness enhancement becomes effective, which retards the
decreasing trends of both the ionization fraction and IGM
temperature. This shows that the DM annihilation may have obvious
contributions to the ionization fraction of the Universe. However,
it seems that the DM component is not able to fully ionize the
Universe as suggested in \cite{2009PhRvD..80c5007B}, in order not to
violate the EGRB observations. It is also shown in
\cite{2009A&A...505..999H} that the model which can fully ionize the
Universe will violate the CMB measurements.

This change of the recombination history will broaden the surface of
last scattering, suppress the temperature fluctuations and enhance
the polarization fluctuations of the CMB. A higher level of residual
ionization fraction will also increase the scattering probability of
CMB photons after decoupling, thus increasing the optical depth of
CMB photons, which has been measured by WMAP
\cite{2009ApJS..180..330K}. The implications for the DM models from
the above effects on the CMB were discussed recently in
\cite{2009A&A...505..999H,2009PhRvD..80d3526S}. Since these effects
are not sensitive to the clumpiness factor $\Delta^2(z)$, we will
not discuss them in detail here. We point out that the mass and
cross section fitted in \cite{2010PhRvD..81b3516L}, and also used in
the current study are consistent with the WMAP constraints at
$2\sigma$ level \cite{2009PhRvD..80d3526S}.

\subsection{21 cm signal from neutral Hydrogen}

The redshifted 21 cm signal from neutral hydrogen can provide
valuable information on the thermal state of gas in the dark ages.
The differential brightness temperature
relative to that of the CMB is given by
\begin{equation}
\delta T_b=16(1-x_{\rm H})(1+\delta)\left(\frac{\Omega_bh}{0.02}\right)
\left[\left(\frac{1+z}{10}\right)
\left(\frac{0.3}{\Omega_m}\right)\right]^{1/2}\left[1-\frac{T_{\rm CMB}
(z)}{T_s}\right] [\rm mK],
\label{eq:deltatb}
\end{equation}
where $x_{\rm H}$ is the hydrogen ionization fraction as defined in
the previous section, $T_{\rm CMB}(z)$ is the CMB temperature at
redshift $z$, $\delta$ is the density contrast of the gas which
should be $0$ for the cosmic mean field\footnote{This formula
neglects the effect from peculiar velocity, the spherical average of
peculiar velocity correction would give a factor $4/3$ in front of
$\delta$ \cite{2008PhRvD..78j3511P}, however, it does not affect our
calculation since we consider the 21 cm signal of the cosmic mean
field, i.e. $\delta=0$.}, $\Omega_m=\Omega_\chi+\Omega_b$, and $T_s$
is the spin temperature of the neutral hydrogen. The spin
temperature, which describes the distribution on the two
hyperfine-split ground states of neutral hydrogen, is given by a
weighted average of the CMB temperature and gas kinetic temperature,
\begin{equation}
T_s=\frac{T_{\rm CMB}(z)+(y_{\alpha,\rm eff}+y_c)T_{\rm igm}}
{1+y_{\alpha,\rm eff}+y_c}.
\end{equation}
The weighting factor depends on the coupling between the spin system
with atoms, which are induced either by atom-atom, atom-proton and
atom-electron collisions ($y_c$), or by resonance scattering of the
Ly$\alpha$ photons ($y_{\alpha,\rm eff}$) which has a color temperature
almost equal to the gas kinetic temperature. Here we adopt $y_c$ from
\cite{2006ApJ...637L...1K} (see also the original references therein).
The effective Ly$\alpha$ efficiency is
\begin{equation}
y_{\alpha,\rm eff}=\frac{P_{10}T_\star}{A_{10}T_{\rm igm}}\times
e^{-0.3(1+z)^{1/2}T_{\rm igm}^{-2/3}}
\left(1+\frac{0.4}{T_{\rm igm}}\right)^{-1},
\end{equation}
in which $T_\star=0.068$ K corresponds to the energy split of the
hyperfine structures of neutral hydrogen, $P_{10}\approx1.3\times10^9
J_\alpha$ with $J_\alpha$ the specific intensity of Ly$\alpha$
radiation, and $A_{10}=2.85\times10^{-15}$ s$^{-1}$ is the Einstein
coefficient of the hyperfine spontaneous transition (see
\cite{2006ApJ...651....1C,2007MNRAS.381.1137C}).

\begin{figure}[!htb]
\begin{center}
\includegraphics[width=0.6\columnwidth]{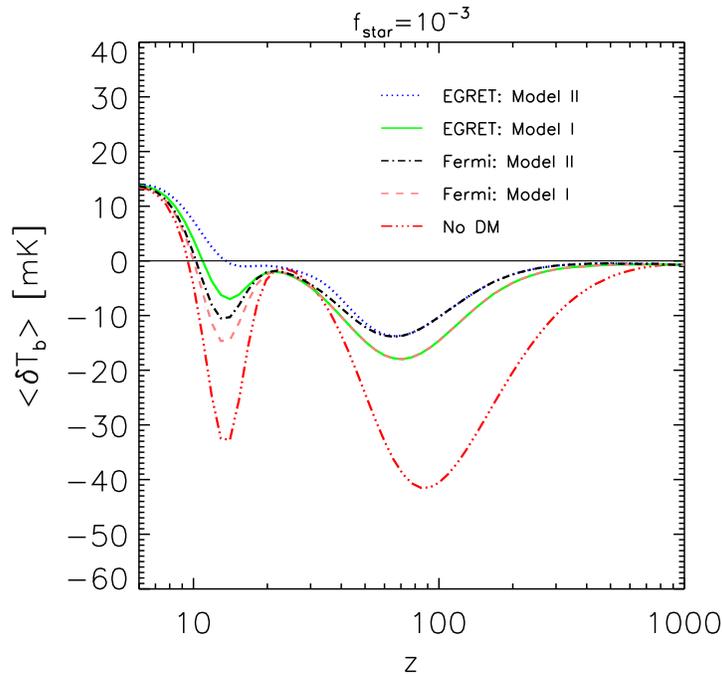}
\caption{Evolution of the 21 cm signal for the four different DM models,
compared with the case without DM annihilation. In all cases, heating
from X-ray and Ly$\alpha$ photons from stars are included, see the text
for details. The thin solid line corresponds to $\delta T_b=0$.}
\label{fig:tb}
\end{center}
\end{figure}

\begin{figure}[!htb]
\begin{center}
\includegraphics[width=0.6\columnwidth]{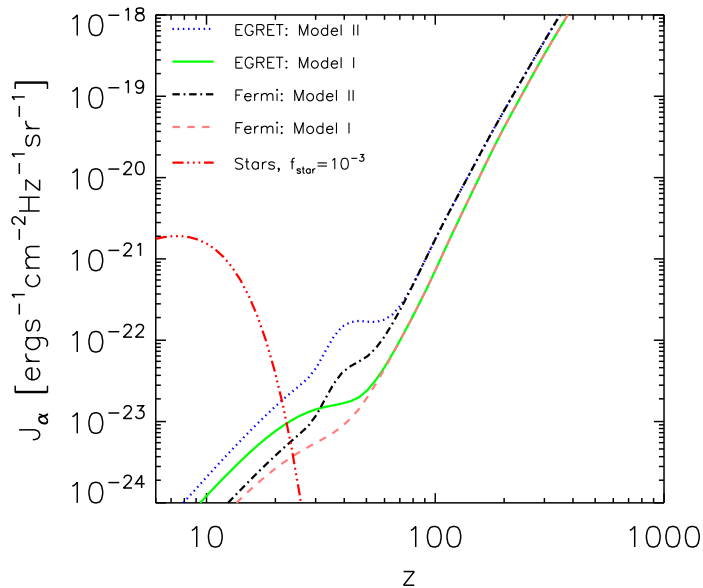}
\caption{Evolution of the DM induced Ly$\alpha$ intensity. For
comparison, we also plot the Ly$\alpha$ intensity from stars in the
same panel. DM annihilation provides an early Ly$\alpha$ background,
but the contribution from stars increases dramatically after
redshift $z\approx$ 30.} \label{fig:Ja}
\end{center}
\end{figure}

We firstly consider the standard picture without the energy
injection from DM annihilation. During the dark age, the scattering
of residual free electrons with CMB photons will keep the gas
temperature comparable with the CMB temperature at the initial
stage. After kinetic decoupling of the baryons from radiation at
$z\sim200$, the gas temperature drops faster, i.e. $T_{\rm
igm}\sim(1+z)^{-2}$. The mean density of the Universe is still high
at this time, and the coupling due to collisions is strong enough to
make $T_s$ almost equal to $T_{\rm igm}$, so we would see an
absorption 21 cm signal (c.f. the line labeled ``No DM'' in
Fig.~\ref{fig:tb}). This absorptional signal peaks at $z \sim 100$.
As the redshift decreases, the density drops and the collisional
coupling becomes less effective, resulting in the spin temperature
approaches $T_{\rm CMB}(z)$ and the 21 cm absorption signal becomes
weak. However, as the stars begin to form, Ly$\alpha$ photons are
produced, and the spin temperature deviates from $T_{\rm CMB}(z)$
and gets close to $T_{\rm igm}$ again. Thus before the IGM is heated
above the CMB temperature, we would see a second absorption peak.
The location of the second peak depends on the model of star
formation, Ly$\alpha$ and X-ray production rate
\cite{2004ApJ...602....1C}. We model the production of the
Ly$\alpha$ background by star formation following
Ref.~\cite{2005ApJ...626....1B}. The evolution of the Ly$\alpha$
background is plotted in Fig.~\ref{fig:Ja}. The contribution to
Ly$\alpha$ from star formation rises sharply after $z\sim40$. This
model is very simple but illustrates qualitatively the main feature
of the 21 cm signal evolution. For our adopted model, the second
absorption peak is at $z\sim 14$. The signal converts from
absorption to emission at $z\sim 10$ due to X-ray heating.

In the above we described the case of the global signal according to
the mean density of gas. In addition, minihalos are formed during the
late dark ages. Within minihalos the temperature of the gas is higher
and the spin-kinetic coupling is also maintained by collisions as the
density is higher. The minihalos could produce a positive 21 cm signal,
but even taking into account the contribution from minihalos, the
overall absorption signal is still apparent
\cite{2002ApJ...572L.123I,2006ApJ...652..849F,2009MNRAS.398.2122Y}.

Now we turn to the cases with DM annihilation. The annihilation of DM
has two effects on the 21 cm signal. On one hand, it heats the IGM up
to a higher temperature and causes a partial ionization, as shown in
Fig.~\ref{fig:xt}. On the other hand, it also produces an early
Ly$\alpha$ background before the formation of the first stars
through de-excitation. The evolutions of $\delta T_b$ for different
DM models as listed in Table \ref{table:halo} are plotted in
Fig. \ref{fig:tb}. The transition redshift at which the IGM temperature
exceeds the CMB temperature is also the redshift at which
the 21 cm signal changes from absorption to emission.

Assuming that a fraction $f_{\alpha}$ of the excitation energy goes into
Ly$\alpha$ photons, the evolution of Ly$\alpha$ intensity is
\begin{equation}
J_\alpha=\frac{c}{4\pi}\frac{[1-x_{\rm H}+f_{\rm He}(1-x_{\rm He})]}
{3(1+f_{\rm He})}\,f_\alpha\,\epsilon_{\rm DM}(z)\frac{n_b}{\nu_\alpha H(z)},
\end{equation}
where $n_b$ is the number density of baryon, $\nu_\alpha=2.47\times
10^{15}$ Hz is the frequency of Ly$\alpha$ photon. We adopt the
energy conversion fraction $f_\alpha=0.5$
\cite{2006PhRvD..74j3502F,2009PhRvD..80d3529N}. The evolution of the
DM annihilation induced Ly$\alpha$ intensity is shown in
Fig.~\ref{fig:Ja} for different models. The DM annihilation produces
a Ly$\alpha$ flux even during the dark ages. As the Universe expands
and the DM density falls, the Ly$\alpha$ flux decreases due to the
drop of annihilation rate. Thanks to the structure formation
$J_\alpha$ shows some enhancement for $z<80$. Due to the expansion
of the Universe, however, the flux which is proportional to the
proper density of the photons still shows a trend of decrease, and
the emission from the stars dominates at lower redshift.

The 21 cm signal with leptonic DM annihilation contribution is shown
in Fig.~\ref{fig:tb}. During the dark age, the absorption signal
would not be as strong as the case without DM annihilation, because
the IGM is heated to higher temperatures. Also the second absorption
peak will become weaker due to the enhancement of DM structures. In
some case there is even no $2nd$ absorption peak. The transition
redshift from absorption to emission will be different from the case
without DM annihilation. However, because there is large uncertainty
of the impact on 21 cm signal from stars, it is not easy to use the
transition redshift to distinguish the model with DM annihilation
from the standard one. We conclude that the magnitude of the first
absorption peak will be more efficient to search for the DM
annihilation signal and constrain the DM models.

%Indeed, it could easily exceed the CMB temperature at redshifts $z>20$,
%so that we would not be able to see the second absorption peak like
%the ``No DM'' case. Because of the saturation of the emission
%signal, the 21 cm signal could not be too strong, which may be bad
%news for observation. Nevertheless, the difference of 21 cm signal
%between DM models and standard (``No DM'') model could in principle
%be distinguished. Also we might see the differences between
%different DM models. Note that the heating of the gas due to DM and
%star formation generally occur at different redshifts, so
%generically we would have two peaks in the 21 cm emission signal.
%The signal decreases at even lower redshift ($z \sim 10$) due to the
%factor $(1+z)^{-1}$ in Eq.~(\ref{eq:deltatb}). At such
%lower redshift much of the intergalactic space would be ionized, and
%in plausible models much of this is due to star formation rather
%than DM annihilation.

\subsection{Suppression on structure formation}

\begin{figure}[!htb]
\begin{center}
\includegraphics[width=\columnwidth]{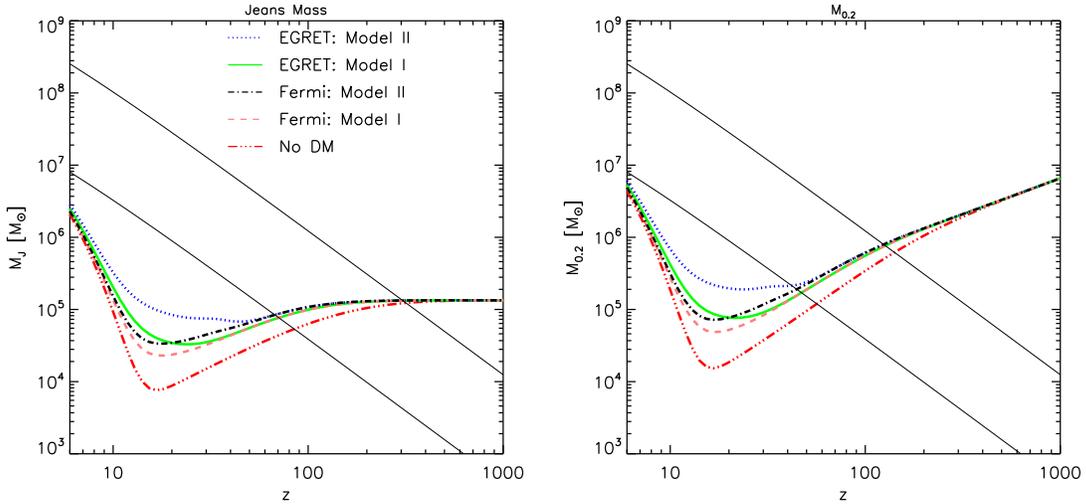}
\caption{Jeans mass ({\it left}) and the minimum mass of a halo that
can hold at least $20\%$ of gas ({\it right}) for different DM
models. The two thin solid lines correspond to the virial mass of
halo with $T_{\rm vir}=10^3$ K ({\it lower}) and $T_{\rm vir}=10^4$
K ({\it upper}) respectively.} \label{fig:mj}
\end{center}
\end{figure}

Heating by DM annihilation can also suppress the accretion of gas by
haloes. The minimum mass for a halo to accrete gas from environment is the
Jeans mass,
\begin{equation}
M_{J}=\frac{\pi}{6}\left(\frac{\pi k_B T_{\rm igm}}{G\mu m_H}\right)
^{3/2}\rho^{-1/2},
\end{equation}
where $\mu$ is the averaged particle weight, $G$ is the gravitational
constant, $k_B$ is Boltzmann constant and $m_H$ is the mass of hydrogen
atom (see \cite{2007MNRAS.375.1399R}). The Jeans mass depends on the
temperature of the gas. In the adiabatic model the temperature of the gas
would decrease as redshift decreases. As a result the Jeans mass would
be lower. However, with the heating of IGM by DM annihilations, the
Jeans mass would instead increase at lower redshift, as shown in the
left panel of Fig.~\ref{fig:mj}.

Without external heating, it is thought that the first stars
(population III or Pop III) might be formed in DM halos with a
virial temperature above $10^3$ K (corresponding to halos with mass
$\sim 10^5$ M$_{\odot}$ at $z \sim 30$). In such halos the trace
amount of $\rm H_2$ in the primordial gas could cool the accreted
gas, which would be bound further and form stars. With the existence
of DM annihilation, the IGM gas will be heated up to higher
temperatures, and the corresponding Jeans mass will also become
higher. However, as shown in Fig.~\ref{fig:mj} the Jeans mass will
be lower than mass with virial temperature $10^3$ K, which means the
impact on the formation of the first stars may be weak.

To account for the effect of DM annihilation on early star
formation, we also follow the prescription given by Oh \& Haiman
\cite{2003MNRAS.346..456O} to estimate the fraction of gas ($f_{\rm
gas}$) left in a halo in the presence of an ``entropy floor'', i.e.
heating of the gas by an external source. This fraction is
determined by $\hat{K}=K_{\rm IGM}/K_o$, where $K_{\rm IGM}=T_{\rm
IGM}(z)/n^{2/3}(z)$ is the entropy of the IGM, and $K_o=T_{\rm
vir}/n^{2/3}(r_{\rm vir})$ is the entropy of a halo at the virial
radius. We interpolate $f_{\rm gas}$ as a function of $\hat {K}$
following \cite{2003MNRAS.346..456O} (see also
\cite{2006ApJ...652..849F}), adopting $\gamma=1$ (i.e. NFW profile).
The {\it right} panel of Fig. \ref{fig:mj} shows the evolution of
the minimum mass of halos that can retain at least $20\%$ of the
gas. This mass is several times larger than the Jeans mass, however,
it is still less than the virial mass supposed to form the first
stars. The DM annihilation is shown not to significantly affect the
early star formation.

\section{Conclusions and Discussions}

In this work we investigate the leptonic DM annihilation, as
indicated by the recent CR lepton data from PAMELA, ATIC, HESS and
Fermi, during the evolution process of the Universe. We mainly focus
on the period after the cosmic recombination. The mass of DM
particle, annihilation cross section and branching ratios to three
lepton flavors are adopted to best-fit the observational CR lepton
data in the Galaxy. Then we calculate the expected EGRB flux from DM
annihilation, including FSR and electron/positron IC emission. The
DM clumpiness effect is included in the calculation according to the
Sheth-Tormen (ST) halo mass function \cite{1999MNRAS.308..119S}. Two
free parameters, the central density slope $\gamma$ and the minimum
halo mass $M_{\rm min}$, are employed to describe the properties of
the DM halo. Even so, the DM halo still suffers the uncertainties of
the concentration. We adopt two kinds of concentration models in
this work, B01 \cite{2001MNRAS.321..559B} and M08
\cite{2008MNRAS.391.1940M}.

The calculated EGRB results are compared with the observational data
from EGRET and Fermi. By requiring the calculated EGRB not to exceed
$2\sigma$ errorbars of the data, we can get a conservative
constraint on parameters $\gamma$ and $M_{\rm min}$. We find that
for the B01 concentration model, the constraints on the model
parameters are relatively weak, while for M08 concentration model in
which the concentration of low mass halo is very large, the
constraints are much stronger. Fermi data can give much stronger
constraints than EGRET data. Even for B01 concentration, Fermi data
can constrain the canonical cold DM prediction of halos, e.g.
$\gamma=1$ and $M_{\rm min}\approx 10^{-6}$ M$_{\odot}$. Since the
spectral shape of the DM contribution is obviously different from
the observational one, we would expect that the DM component does
not contribute significantly to the EGRB. Therefore taking into
account the background contribution we can expect that the remaining
possible DM can be much reduced. We suggest that a warm DM scenario
with non-thermal production of DM particles
\cite{2001PhRvL..86..954L} might be consistent with the data safely.

Then we investigate the effects of such models of DM annihilation on
the reionization and heating of the IGM. In our model, in the
absence of DM annihilation but with X-ray only, the ionization
fraction and the IGM temperature till $z\approx 6$ are $O(10^{-3})$
and $O(100)$ K respectively\footnote{Note that the ionization
fraction seems to be lower than that derived according to
observations of Ly$\alpha$ forest and the WMAP measurement of the
electron scattering optical depth \cite{2009arXiv0908.3891P}. In
this work we only adopt a simple description of the ordinary
reionization precess and do not intend to give good description to
Ly$\alpha$ forest and WMAP data. In fact, there are still large
uncertainties on the meansurements of mean transmittance of
Ly$\alpha$ flux, especially at high redshifts. A detailed model of
the reionization history based on the Ly$\alpha$ and WMAP
observations is strongly model dependent and out of the purpose of
this study.}. If the DM annihilation is taken into account as an
energy source, the cosmic ionization fraction will have a higher
plateau (of the order $O(10^{-3})$) during the evolution history.
Also the IGM temperature will be higher by several times for
redshift $10<z<100$. However, the effects of DM clumpiness on the
enhancement of the energy deposition rate are not effective. In our
treatment the energy deposition process within the evolution frame
of the Universe is taken into account, which will result in a much
lower energy deposition rate compared with the case with
instantaneous energy deposition. We further find that it is
impossible for the DM models considered here to fully ionize the
Universe without conflicting with the EGRB data.

Finally we discuss the influence of the DM annihilation on the
neutral hydrogen 21 cm signal and the suppression on structure
formation, which may connect with observations more directly than
the ionization fraction and the IGM temperature. It is found that
the global 21 cm signal for the models with DM annihilation differs
somewhat from the ``standard'' scenario without DM annihilation. Due
to the rise of the IGM temperature during the dark age, the 21 cm
absorption signal is much smaller than the ``standard'' case. Also,
the Ly$\alpha$ flux supplied by the DM annihilation would make the
spin temperature be closer to the gas kinetic temperature at $z<40$.
But again due to the heating, the gas kinetic temperature could
exceed CMB temperature earlier than the ``standard'' case. In some
model the $2nd$ absorption peak with redshift between $10$ and $20$
in the ``standard'' scenario may not appear. Future 21 cm
observations may have the potential to test this leptonic DM
annihilation scenario and constrain the model parameters, despite
the technical difficulty in practice.

The heating of the IGM will also affect the Jeans mass, which
characterizes the critical mass of DM halo that can accrete gas. We
show that with the constraint from EGRB observations, the leptonic
DM annihilation scenarios are not able to affect the structure
formation process significantly.

\appendix*

\section{Clumpiness enhancement factor of DM annihilation}

The density profile of a DM halo with mass $M_{\rm vir}$ can be
parameterized as
\begin{equation}
\rho(r)=\frac{\rho_s}{(r/r_s)^{\gamma}(1+r/r_s)^{3-\gamma}}=
\rho_s\times \tilde{\rho}(x\equiv r/r_s),
\end{equation}
here we adopt a free parameter $\gamma$ to represent the inner profile
of DM distribution inside a halo. The scale radius $r_s$ is determined
by the concentration parameter $c_{\rm vir}$ of the halo
\begin{equation}
r_s=\frac{r_{\rm vir}}{c_{\rm vir}(2-\gamma)},
\end{equation}
where
\begin{equation}
r_{\rm vir}=\left(\frac{3M_{\rm vir}}{4\pi\Delta_{\rm vir}(z)\rho_\chi(z)}
\right)^{1/3},
\end{equation}
is the virial radius, with $\rho_\chi(z)=\rho_\chi(1+z)^3$ the DM density
at redshift $z$, $\Delta_{\rm vir}(z)=(18\pi^2+82y-39y^2)/(1+y)$,
$y=\Omega_m(z)-1$ and $\Omega_m(z)=\frac{\Omega_m(1+z)^3}{\Omega_m(1+z)^3+
\Omega_\Lambda}$ \cite{1998ApJ...495...80B}. The concentration parameter
as a function of halo mass $M_{\rm vir}$ can be extracted from N-body
simulations. For the low mass halos which are beyond the
resolution of simulations, the concentration is obtained by
extrapolation.In this paper we considered two extrapolation
models. One is the analytical model developed in
Ref. \cite{2001MNRAS.321..559B} (B01), and the other is the direct
extrapolation of the fitting results from simulations under the WMAP5
cosmology \cite{2008MNRAS.391.1940M} (M08). For both models we
employ the redshift evolution $c_{\rm vir}(z)=c_{\rm vir}(z=0)/(1+z)$
\cite{2001MNRAS.321..559B}. The scale density $\rho_s$ is determined by
normalizing the mass of the halo to $M_{\rm vir}$.

The annihilation $\gamma$-ray intensity from a single halo is
\begin{eqnarray}
L(M_{\rm vir})&=& 4\pi\int_0^{r_{\rm vir}}\rho^2(r)r^2{\rm d}r \nonumber\\
&=& 4\pi r_s^3\rho_s^2 \int_0^{x_{\rm max}}\tilde{\rho}^2(x)x^2{\rm d}x
\nonumber\\
&=& \frac{M_{\rm vir}\Delta_{\rm vir}(z)\rho_\chi(z)x_{\rm max}^3}{3}
\frac{\int\tilde{\rho}^2(x)x^2{\rm d}x}{\left(\int\tilde{\rho}(x)x^2
{\rm d}x\right)^2},
\label{Lsingle}
\end{eqnarray}
where $x_{\rm max}=c_{\rm vir}(2-\gamma)$.
For a population of DM halos with comoving number density distribution
${\rm d}n(z)/{\rm d}M_{\rm vir}$ the total annihilation luminosity is
\begin{equation}
L_{\rm tot}=\int{\rm d}M_{\rm vir}\frac{{\rm d}n(z)}{{\rm d}M_{\rm vir}}
(1+z)^3L(M_{\rm vir})=\rho_\chi^2(1+z)^6\times \Delta^2(z),
\end{equation}
in which one of the $(1+z)^3$ factor comes from $\rho_\chi(z)$ term in
Eq.(\ref{Lsingle}), and the other $(1+z)^3$ converts the comoving halo
mass function to the physical one. In the above equation we define the
clumpiness enhancement factor as
\begin{equation}
\Delta^2(z)=\frac{\Delta_{\rm vir}(z)}{3\rho_\chi}\int{\rm d}M_{\rm vir}
M_{\rm vir}\frac{{\rm d}n(z)}{{\rm d}M_{\rm vir}} \frac{\int\tilde
{\rho}^2(x)x^2{\rm d}x}{\left(\int\tilde{\rho}(x)x^2{\rm d}x\right)^2}
x_{\rm max}^3.
\end{equation}
This is used in Eq.~(\ref{phi}).

Finally we specify the halo mass function. The comoving number density
distribution of DM haloes can be expressed as
\begin{equation}
\frac{{\rm d}n(z)}{{\rm d}M_{\rm vir}}=\frac{\rho_\chi}{M_{\rm vir}}
\sqrt{\frac{2A^2a^2}{\pi}}\left[1+(a\nu^2)^{-p}\right]\exp\left(-a\nu^2/2
\right)\frac{{\rm d}\nu}{{\rm d}M_{\rm vir}},
\end{equation}
where $\nu=\delta_c(z)/\sigma(M_{\rm vir})$, $\delta_c(z)=1.68/D(z)$
is the critical overdensity in spherical collapse model, $D(z)$ is the
linear growth factor \cite{1992ARA&A..30..499C}. $A$, $a$ and $p$ are
constants. For $(A,\,a,\,p)=(0.5,\,1,\,0)$ it is the Press-Schechter (PS)
formula \cite{1974ApJ...187..425P}, and for $(A,\,a,\,p)=(0.322,\,0.707,
\,0.3)$ it is Sheth-Tormen (ST) formula \cite{1999MNRAS.308..119S}.
In this work we adopt the ST mass function.
$\sigma^2(M_{\rm vir})$ is the average variance of density field
\begin{equation}
\sigma^2(M_{\rm vir})=\frac{1}{2\pi^2}\int W^2(kR_{\rm M})P_\delta(k)
k^2{\rm d}k,
\end{equation}
with the top-hat window function $W(x)=3(\sin{x}-x\cos{x})/x^3$.
$R_{\rm M}=\left(3M_{\rm vir}/4\pi\rho_m\right)^{1/3}$ is a radius
within which a mass $M_{\rm vir}$ is contained with a uniform matter
density field. The matter power spectrum $P_\delta(k)$ is expressed as
\begin{equation}
P_\delta(k)=A_s(k\cdot{\rm Mpc})^{n_s}T^2(k),
\end{equation}
where $A_s$ is normalized using $\sigma_8$, transfer function $T(k)$
is obtained from a fit of CDM model \cite{1986ApJ...304...15B},
\begin{equation}
T(q)=\frac{\ln(1+2.34q)}{2.34q}\left[1+3.89q+(16.1q)^2+(5.46q)^3+
(6.71q)^4\right]^{-0.25}
\end{equation}
with $q=k/h\Gamma$ and $\Gamma=\Omega_m h\exp[-\Omega_b(1+\sqrt{2h}/
\Omega_m)]$.

\acknowledgments

We are grateful to the anonymous referee for the useful comments and 
suggestions. We thank Feng Huang, Yupeng Yang, Xiaoyuan Huang and 
Yidong Xu for helpful discussions. This work is supported in part by 
the Natural Sciences Foundation of China (No. 10773011, No. 10525314, 
10533010), by the Chinese Academy of Sciences under the grant No. 
KJCX3-SYW-N2, and by the Ministry of Science and Technology National 
Basic Science Program (Project 973) under grant No. 2007CB815401.

%\bibliography{refs}
\bibliography{/home/yuanq/work/cygnus/tex/refs}
\bibliographystyle{apsrev}

\end{document}